\documentclass[10pt]{article}

\usepackage[authoryear,round]{natbib}                   
\usepackage{amsmath}                                                        
\usepackage{graphicx}                                                       
\usepackage{subfigure}
\usepackage{enumitem}                                                    
\usepackage{hyperref}
\usepackage{amssymb}                                                        
\usepackage[mathscr]{eucal}                                             
\usepackage{cancel}                                                             
\usepackage[normalem]{ulem}                                                                 
\usepackage{pstricks}
\usepackage{rotating}
\usepackage{lscape}
\usepackage[paperwidth=8.5in,paperheight=11in,top=1.25in, bottom=1.25in, left=1.00in, right=1.00in]{geometry}
\usepackage{mathtools}                                                      
\mathtoolsset{showonlyrefs=true}                                    
\usepackage{fixltx2e,amsmath}                                           
\MakeRobust{\eqref}

\usepackage{mathdots}
\usepackage{amsthm}                                                             
\allowdisplaybreaks                                                             
\theoremstyle{plain}
\newtheorem{theorem}{Theorem}

\theoremstyle{definition}

\newtheorem{remark}[theorem]{Remark}
\newtheorem{assumption}[theorem]{Assumption}

\def \s {{\sigma}}

\newcommand{\<}{\langle}
\renewcommand{\>}{\rangle}
\renewcommand{\(}{\left(}
\renewcommand{\)}{\right)}
\renewcommand{\[}{\left[}
\renewcommand{\]}{\right]}

\newcommand\Eb{\mathbb{E}}

\newcommand\Qb{\mathbb{Q}}
\newcommand\Rb{\mathbb{R}}
\newcommand\Ib{\mathbb{I}}

\newcommand\Ac{\mathscr{A}}

\newcommand\Ec{\mathscr{E}}
\newcommand\Fc{\mathscr{F}}
\newcommand\Gc{\mathscr{G}}

\newcommand\Lc{\mathscr{L}}
\newcommand\Mc{\mathscr{M}}
\newcommand\Nc{\mathscr{N}}
\newcommand\Oc{\mathscr{O}}

\newcommand\om{\omega}
\newcommand\Om{\Omega}
\newcommand\sig{\sigma}

\newcommand\gam{\gamma}
\newcommand\Gam{\Gamma}

\newcommand\del{\delta}

\newcommand\xb{\bar{x}}
\newcommand\yb{\bar{y}}

\newcommand\Hv{\mathbf{H}}
\newcommand\Cv{\mathbf{C}}
\newcommand\mv{\mathbf{m}}
\newcommand\etav{\boldsymbol\eta}

\newcommand\At{\widetilde{\Ac}}

\newcommand\Xt{\widetilde{X}}
\newcommand\xt{\widetilde{x}}
\newcommand\ut{\widetilde{u}}

\def \p {{\partial}}

\renewcommand\d{\partial}

\newcommand\dd{\mathrm{d}}
\newcommand\ee{\mathrm{e}}
\newcommand\BS{\text{\rm BS}}

\begin{document}

\title{
A Taylor series approach to pricing and implied vol for LSV models
}

\author{
Matthew Lorig
\thanks{ORFE Department, Princeton University, Princeton, United States.
\textbf{E-mail:} \url{mlorig@princeton.edu}.
Work partially supported by NSF grant DMS-0739195.}
\and
Stefano Pagliarani
\thanks{Dipartimento di Matematica, Universit\`a di Padova, Padova, Italy.
\textbf{E-mail:} \url{stefanop@math.unipd.it}.}
\and
Andrea Pascucci
\thanks{Dipartimento di Matematica,
Universit\`a di Bologna, Bologna, Italy.
\textbf{E-mail:} \url{andrea.pascucci@unibo.it}.}
}

\date{This version: \today}

\maketitle

\begin{abstract}
Using classical Taylor series techniques, we develop a unified approach to pricing and implied volatility for European-style options in a general local-stochastic volatility setting.  Our price approximations require only a normal CDF and our implied volatility approximations are fully explicit (ie, they require no special functions, no infinite series and no numerical integration).  As such, approximate prices can be computed as efficiently as Black-Scholes prices, and approximate implied volatilities can be computed nearly instantaneously.
\end{abstract}

%
%

\section{Introduction}
There are a myriad of local volatility (LV), stochastic volatility (SV) and local-stochastic volatility (LSV) models for which European option prices can be computed explicitly (e.g., CEV, JDCEV, Heston, three-halves, SABR with zero correlation).  However, these explicit formulas require special functions, a large number of terms, or numerically integrating highly oscillatory functions.  As such, computing option prices with these formulas can be both delicate and computationally expensive.  Moreover, when time-dependent parameters are introduced, which are needed to fit the term-structure of implied volatility, many of these models lose the analytic tractability that made them desirable in the first place.
\par
For the purposes of calibration, one requires implied volatilities rather than prices.  And, there are a plethora of explicit implied volatility approximations for LV, SV and LSV models which are useful in this respect (e.g., CEV, Heston, SABR, $\lambda$-SABR).  However, these expansions rely on specific model dynamics, which may not be appropriate for a given underlying.  And, once again, introducing explicit time-dependence can be problematic.
\par
In this article, we introduce a simple yet effective method for computing approximate European option prices and corresponding implied volatilities for any SV, LV or LSV model with time-dependent drift, diffusion and killing coefficients (we allow for the possibility of default).  Our method, which is based on the classical Taylor series expansion, results in pricing approximations that require only a normal CDF and implied volatility approximations that are fully explicit (ie, they do not require any numerical integration nor do they require special functions).  As such, approximate European option prices can be computed as efficiently as in the Black-Scholes model and implied volatilities can be computed nearly instantaneously.

%
%

\section{General local-stochastic volatility models with default}
\label{sec:model} For simplicity, we assume a frictionless market, no arbitrage, zero interest
rates and no dividends.  All our results can be easily extended to deterministic interest rates.
We take, as given, an equivalent martingale measure $\Qb$, chosen by the
market on a complete filtered probability space $(\Om,\Fc,\{\Fc_t,t\geq0\},\Qb)$.  The filtration
$\{\Fc_t,t\geq0\}$ represents the history of the market.  All stochastic processes defined below
live on this probability space and all expectations are taken with respect to $\Qb$.  We consider
an asset $S$ whose risk-neutral dynamics are given by
\begin{align}
\begin{aligned}
S_t
    &=  \Ib_{\{ \zeta > t\}} \ee^{ X_t } , \\
\dd X_t
    &=  \mu(t,X_t,Y_t) \dd t + \sig(t,X_t,Y_t) \dd W_t , & X_0 &= x \in \Rb , \\
\dd Y_t
    &=  \alpha(t,X_t,Y_t) \dd t + \beta(t,X_t,Y_t) \dd B_t , & Y_0 &= y \in \Rb, \\
\dd \< W, B \>_t
    &=  \rho(t,X_t,Y_t) \, \dd t , &
|\rho|
    &< 1 ,
\end{aligned} \label{eq:StochVol}
\end{align}
where $\zeta$ is a stopping time which represents a possible default event
\begin{align}
\zeta
    &=  \inf \big\{ t \geq 0 : \int_0^t \gam(s,X_s,Y_s) ds \geq \Ec \big\} ,
\end{align}
with $\Ec$ exponentially distributed and independent of $X$. As the asset price $S$ must be a
martingale, the drift function $\mu$ must be given by
\begin{align}
\mu(t,x,y)
    &=  - \frac{1}{2} \sig^2(t,x,y) + \gam(t,x,y).
\end{align}
Equation \eqref{eq:StochVol} includes virtually all local volatility models, all one-factor
stochastic volatility models, and all one-factor local-stochastic volatility models.  Moreover,
the results of this paper can be extended in a straightforward fashion to include models with $n$
non-local factors of volatility.  Though, for simplicity, we restrict our analysis to a single
non-local factor.

Denote by $V$ the no-arbitrage price of European derivative expiring at time $T$ with payoff
$H(S_T)$. It is well known (see, for instance, \cite{yorbook}) that
\begin{align}\label{e1}
V_t
    &=  K + \Ib_{\{\zeta>t\}} \Eb \[\ee^{-\int_t^T \gam(s,X_s,Y_s) \dd s} \(h(X_T)-K \)  | X_t , Y_t
    \], &
t
    &<T,
\end{align}
where $K:=H(0)$ and $h(x):=H(\ee^x)$.
Then, to value a European-style option, one must compute functions of the form
\begin{align}
u(t,x,y)
    &:= \Eb \[\ee^{- \int_t^T \gam(s,X_s,Y_s) \dd s}  h(X_T) \mid X_t = x , Y_t = y \] . \label{eq:v}
\end{align}
The function $u$, defined by \eqref{eq:v}, satisfies the Kolmogorov backward equation
\begin{align}
(\d_t + \Ac) u
    &=  0 , &
u(T,x,y)
    &=  h(x,y) , \label{eq:v.pde}
\end{align}
where the operator $\Ac$ is given explicitly by
\begin{align}
\Ac
    &=  a(t,x,y) ( \d_x^2 - \d_x ) + \alpha(t,x,y) \d_y + b(t,x,y) \d_y^2 + c(t,x,y) \d_x \d_y + \gam(t,x,y)(\d_x - 1), \label{eq:A}
\end{align}
and where the functions $a$, $b$ and $c$ are defined as
\begin{align}
a(t,x,y)
    &:=  \frac{1}{2}\sig^2(t,x,y) , &
b(t,x,y)
    &:=  \frac{1}{2}\beta^2(t,x,y) , &
c(t,x,y)
    &:=  \rho(t,x,y) \sig(t,x,y) \beta(t,x,y) . \label{eq:abc}
\end{align}
\begin{remark}[Deterministic interest rates]
For deterministic interest rates $r(t)$ one must compute expectations of the form
\begin{align}
\ut(t,\xt,y)
    &:= \Eb \[\ee^{- \int_t^T r(s) + \gam(s,\Xt_s,Y_s) \dd s}  h(\Xt_T) | \Xt_t = \xt , Y_t = y \] , &
    &\text{where}&
\dd \Xt_t
    &= \dd X_t + r(t) \dd t.
\end{align}
In this case a simple change of variables
\begin{align}
u(t,x(t,\xt),y)
    &:= \ee^{\int_t^T r(s)} \ut(t,\xt,y)  , &
x(t,\xt)
    &:= \xt + \int_t^T r(s) \dd s, \label{eq:new.u}
\end{align}
reveals that $u$ defined as in \eqref{eq:new.u}, satisfies \eqref{eq:v.pde}.
\end{remark}

%
%

\section{Pricing approximation: a Taylor series approach}
\label{sec:approximating}
To construct an approximate solution of Cauchy problem \eqref{eq:v.pde}, we push forward the ideas
in \cite{lorig-pagliarani-pascucci-2,Pascucci2011}, which are based on expanding the symbol of
L\'evy-type integro-differential pricing operators. Our goal is to introduce an unified approach to pricing and implied volatility based on the classical Taylor series approximation. Specifically, for any analytic function $f=f(t,x,y)$, we can expand about a fixed point $(\xb,\yb) \in \Rb^2$ as follows
\begin{align}
f(t,x,y)
    &= \sum_{n=0}^\infty \sum_{h=0}^n f_{n-h,h}(t) (x-\xb)^{n-h} (y-\yb)^h , &
f_{n-h,h}(t)
    &:= \frac{1}{(n-h)!\,h!}\d_x^{n-h} \d_y^h f(t,\xb,\yb) .
\end{align}
For brevity, when $n=h=0$ we shall simply write $f_{0}$ instead of $f_{0,0}$.
Applying this idea to the coefficients ($a$, $\alpha$, $b$, $c$, $\gam$) we find that, formally, the operator $\Ac$ in \eqref{eq:A} admits an expansion of the form
\begin{align}
\Ac
    &=  \sum_{n=0}^\infty \Ac_n, &
\Ac_n
        &:=  \sum_{h=0}^n  (x-\xb)^{n-h} (y-\yb)^h \Ac_{n-h,h}, \label{eq:A.expand}
\end{align}
where $\{\Ac_{n-h,h}\}$ is a sequence of differential operators with time-dependent coefficients
\begin{align}
\Ac_{n-h,h}
    &:=  a_{n-h,h}(t) ( \d_x^2 - \d_x ) + \alpha_{n-h,h}(t) \d_y + b_{n-h,h}(t) \d_y^2
                + c_{n-h,h}(t) \d_x \d_y + \gam_{n-h,h}(t)(\d_x - 1) . \label{eq:A.nk}
\end{align}
We shall assume henceforth that the operator $\Ac$ is parabolic, which is typically the case in financial applications.
In light of the above expansion for $\Ac$, we also expand the pricing function $u$ as follows
\begin{align}
  u  &=  \sum_{n=0}^\infty u_n . \label{eq:v.expand}
\end{align}
Inserting \eqref{eq:A.expand} and \eqref{eq:v.expand} into \eqref{eq:v.pde} we find that the functions $\{u_{n}\}$ satisfy the following sequence of Cauchy problems
\begin{align}
(\d_t + \Ac_0) u_0
    &=  0, &
u_0(T,x,y)
    &=  h(x,y), \label{eq:v.0.pide} \\
(\d_t + \Ac_0 ) u_n
    &= - \sum_{k=1}^{n} \Ac_k u_{n-k}, &
u_n(T,x,y)
    &=  0 . \label{eq:v.n.pide}
\end{align}
As we show below, one can find an explicit expression for the $n$th function $u_n$ by using only general properties of distribution functions such as the classical Chapman-Kolmogorov equation and the standard Duhamel's principle.
\par
First, consider the Cauchy problem \eqref{eq:v.0.pide}. The operator $\Ac_0$ is a parabolic operator with time-dependent coefficients.   Thus, the solution $u_0$ can be written as
\begin{align}\label{e10}
u_0(t,x,y)
    &=  \ee^{-\int_t^T \gamma_{0}(s) \dd s} \int_{\mathbb{R}^2} \Gamma_0(t,x,y;T,\xi,\omega) h(\xi,\omega)\, \dd\xi \, \dd\om,
\end{align}
where $\Gamma_0(t,x,y;T,\xi,\omega)$ is a two-dimensional Gaussian density
\begin{align}
\Gamma_0(t,x,y;T,\xi,\omega)
    &=  \frac{1}{ 2\pi \sqrt{|\Cv|} } \exp\( -\frac{1}{2}(\etav - \mv)^\text{T} \Cv^{-1} (\etav - \mv) \) , &
\etav
    &=  \begin{pmatrix}
            \xi \\ \om
       \end{pmatrix} , \label{e22}
\end{align}
with covariance matrix $\Cv$ and mean vector $\mv$ given by:
\begin{align}
\Cv
    &=  \begin{pmatrix}
            2 \int_t^T a_{0}(s) \dd s & \int_t^T c_{0}(s) \dd s  \\
            \int_t^T c_{0}(s) \dd s & 2\int_t^T b _{0}(s) \dd s
            \end{pmatrix} , &
\mv
    &=  \begin{pmatrix}
            x+\int_t^T \left(\gamma_{0}(s)-a_{0}(s)\right) \dd s \\ y+ \int_t^T \alpha_{0}(s) \dd s
         \end{pmatrix} .
\end{align}
\par
Next we consider Cauchy problem \eqref{eq:v.n.pide} with $n=1$.  Let $h=\delta_{(X,Y)}$ so that $u$ corresponds directly to the transition density of the process, $u(t,x,y)=\Gamma(t,x,y;T,X,Y)$.  For any operator $\Ac$, let us denote by $\At$ its formal adjoint, which is obtained by integrating by parts.  For clarity, in the computations below we write $\Ac=\Ac^{(x,y)}(t)$  to indicate that $\Ac$ takes $t$ as an argument and acts on the variables $(x,y)$.  We have
\begin{align}
&u_1(t,x,y) \ee^{\int_t^T \gamma_{0}(s) \dd s} \\
&=  \int_{t}^{T} \dd s \int_{\mathbb{R}^2} \dd\xi \, \dd\om \,
        \Gamma_0(t,x,y;s,\xi,\omega) \Ac_1^{(\xi,\omega)}(s) \Gamma_0(s,\xi,\omega;T,X,Y) &
        &\text{(by \eqref{eq:v.n.pide} and Duhamel's princple)} \\
&=  \int_{t}^{T} \dd s \int_{\mathbb{R}^2} \dd\xi \, \dd\om \,
        \( \At_{1}^{(\xi,\om)}(s) \Gamma_0(t,x,y;s,\xi,\omega) \) \Gamma_0(s,\xi,\om;T,X,Y) &
        &\text{(from integration by parts)} \\
&=  \int_{t}^{T} \dd s \, \Gc_{1}^{(x,y)}(t,s) \int_{\mathbb{R}^2} \dd\xi \, \dd\om \,
        \Gamma_0(t,x,y;s,\xi,\om) \Gamma_0(s,\xi,\om;T,X,Y) &
        &\text{(see equations \eqref{e32} and \eqref{eq:identity})} \\
&=  \int_{t}^{T} \dd s \, \Gc_{1}^{(x,y)}(t,s)
        \Gamma_0(t,x,y;T,X,Y) &
        &\text{(by Chapman-Kolmogorov)}
\end{align}
Multiplying both sides by $\ee^{-\int_t^T \gamma_{0}(s) \dd s}$ and using \eqref{e10} we find
\begin{align}
u_1(t,x,y)
    &=  \Lc_1 u_0(t,x,y), &
\Lc_1
    &:= \int_{t}^{T} \dd s \, \Gc_{1}(t,s) , \label{eq:L1}
\end{align}
where, now it is understood that the operator $\Lc_1$ takes $t$ as an argument and acts on the variables $x,y$.
For higher orders, using results from Appendix \ref{sec:gaussian} we find
\begin{align}\label{eq:un}
  u_n(t,x,y)   &=  \Lc_n u_0(t,x,y) , & \Lc_n
    &=  \sum_{h=1}^n \int_t^T \dd s_1 \cdots \int_{s_{h-1}}^T \dd s_h
      \sum_{i\in I_{n,h}}\Gc_{i_{1}}(t,s_1) \cdots \Gc_{i_{h}}(t,s_h) ,
\end{align}
where\footnote{ For instance, for $n=3$ we have $I_{3,3}=\{(1,1,1)\}$, $I_{3,2}=\{(1,2),(2,1)\}$
and $I_{3,1}=\{(3)\}$. }
\begin{align}
I_{n,h}
    &=  \{i=(i_{1},\dots,i_{h})\in\mathbb{N}^{h} \mid i_{1}+\dots+i_{h}=n\} , &
1
    & \le h \le n ,
\end{align}
and $\Gc_{n}(t,s)$ is an operator
\begin{align}
\Gc_{n}(t,s)
    &:= \sum_{h=0}^n \Mc_{n-h,h}(t,s) \Ac_{n-h,h}(s) , &
\Mc_{h,k}(t,s)
    &:= \( \Mc_1(t,s) \)^h \( \Mc_2(t,s) \)^k , \label{e32}
\end{align}
with
\begin{align}
\Mc_1(t,s)
    &:= (x-\bar{x}) + \int_t^s \left(\gamma _{0}(q)-a_{0}(q)\right) \dd q
            +2 \int_t^s a_{0}(q) \dd q \, \p_x+ \int_t^s c_{0}(q) \dd q \, \p_y , \label{eq:M1} \\
\Mc_2(t,s)
    &:= (y-\bar{y}) + \int_t^s \alpha _{0}(q) \dd q + \int_t^s c_{0}(q) \dd q\, \p_x
            +2 \int_t^s b_{0}(q) \dd q\, \p_y . \label{eq:M2}
\end{align}
An equivalent representation for $\Lc_n$ is given in \cite[Theorem 9]{lorig-pagliarani-pascucci-2} for the time-independent undefaultable case.
\begin{remark}[Accuracy of the pricing approximation]
Asymptotic convergence results were proved in \cite{pascucci,lorig-pagliarani-pascucci-1}.  Precisely, assume that the functions $a=a(t,x,y)$, $\alpha=\alpha(t,x,y)$, $b=b(t,x,y)$ and $c=c(t,x,y)$ are differentiable up to order $n$ with bounded and Lipschitz continuous derivatives. Assume also that the covariance matrix is uniformly positive definite and bounded.
Let $(\xb,\yb)=(x,y)$.  Then for any $N\in\mathbb{N}$ we have
\begin{align}
 u(t,x,y)=\sum_{n=0}^N u_n(t,x,y)+\Oc \left((T-t)^{\frac{N+1}{2}}\right)\qquad \text{\rm as } t\to T^{-}. \label{eq:accuracy}
\end{align}
\end{remark}
\begin{remark}[Practical implementation]
Notice that after a few terms the expression for $\Lc_n$ becomes very long.  In practice, the formulas are feasible only for $n \le 4$.  However, in light of \eqref{eq:accuracy}, it is sufficient to get very accurate results with $n=2$ or $n=3$.
\end{remark}
\begin{remark}[Numerical efficiency]
When an option payoff is a function of $x$ only (which is typically the case), then computing the terms in the option price expansion require no integration and no special functions other than a one-dimensional normal CDF.  As such, the pricing approximation is as efficient to compute as the Black-Scholes price. Moreover, in the case of (possibly defaultable) bonds, approximate prices are fully explicit; no integration or special functions are required.
\end{remark}

%
%

\section{Implied volatility: a Taylor series approach}
\label{sec:impvol}
European Call and Put prices are commonly quoted in units of implied volatility rather than in
units of currency.  In fact, in the financial industry, model parameters for the risk-neutral
dynamics of a security are routinely obtained by calibrating to the market's implied volatility
surface.  Because calibration requires computing implied volatilities across a range of strikes
and maturities and over a large set of model parameters, it is extremely useful to have a method
of computing implied volatilities quickly.
\begin{assumption}
In \emph{this section only}, we assume $\gam(t,x,y)=0$ (ie, no default).
\end{assumption}
For fixed $(t,T,x,k)$, denote by $u^{\BS}(\s)$
the \emph{Black-Scholes price of a Call option} considered as a function of the volatility
\begin{align}
u^{\BS}(\sig)
    &:= \ee^x \Nc(d_{+}(\sig)) - \ee^{k} \Nc(d_{-}(\sig)) , &
d_{\pm}(\sig)
    &:= \frac{1}{\sig \sqrt{T-t}} \( x - k \pm \frac{\sig^2 }{2}(T-t)  \) , \label{eq:BS}
\end{align}
where $\Nc$ is the CDF of a standard normal random variable.  The \emph{implied volatility} corresponding to a Call
price $u\in\,((\ee^{x}-\ee^{k})^+,\ee^x)$ is defined as the unique strictly positive real solution
$\sig$ of the equation
\begin{align}
 u^{\BS}(\sig)    &= u.   \label{eq:imp.vol.def}
\end{align}
Our goal is to find the implied volatility $\sig$ that corresponds to our price expansion $u = \sum_{n=0}^\infty u_n$.  To this end, we assume that $\sig$ has an expansion of the form
\begin{align}
\sig
    &=  \sig_0 + \del , &
\del
    &=  \sum_{n=1}^\infty \sig_n . \label{eq:sigma.expand}
\end{align}
To find the unknown terms in the sequence $\{ \sig_n \}$ we simply insert \eqref{eq:v.expand} and \eqref{eq:sigma.expand} into \eqref{eq:imp.vol.def} and expand $u^{\BS}(\sig_0 + \del)$ in a Taylor series about the point $\sig_0$, ie,
\begin{align}
u^{\BS}(\sig_0 + \del)
    &=  \sum_{n = 0}^\infty \frac{\del^n}{n!} \d_\sig^n u^\BS(\sig_0)
    =       \sum_{n = 0}^\infty u_n .
\end{align}
From the above equation, one can find the unknown terms in the sequence $\{ \sig_n \}$
iteratively.  The explicit expressions are obtained in Theorem $15$ of
\cite{lorig-pagliarani-pascucci-2} and Theorem 4.3 of \cite{lorigCEV}. 
We have
\begin{align}
\sig_0
    &=  \sqrt{\frac{2\int_t^T a_{0}(s) \dd s}{T-t}} , \label{eq:sig.0} \\
 \sig_{n}
    &=  \frac{u_{n}}{\d_\sig u^\BS(\sig_0)}-\frac{1}{n!}\sum_{h=2}^{n}\frac{\d^{n}_{\sig}u^{\BS}(\sig_{0})}{\d_{\sig}u^{\BS}(\sig_{0})}
        \mathbf{B}_{n,h}\big(\sig_{1},2!\sig_{2},3!\sig_{1},\dots,(n-h+1)!\sig_{n-h+1}\big), \label{eq:bell}
\end{align}
where $\mathbf{B}_{n,h}$ denotes the $(n,h)$-th partial Bell\footnote{Partial Bell polynomials are
implemented in Mathematica as $\mathrm{BellY[n,h,\{x_{1},\dots,x_{n-h+1}\}]}$.} polynomial.  Note, in finding $\sig_0$ we used the fact that the leading term in the price expansion $u_0$ is simply $u^{\BS}(\sig_0)$ with $\sig_0$ as defined in \eqref{eq:sig.0}.  Explicitly, the first three terms in \eqref{eq:bell} are
\begin{align}\label{eq:sigmas}
\sig_1
    &=  \frac{u_1}{\d_\sig u^{\text{\rm BS}}\left(\sig_0\right)}, &
\sig_2
    &=  \frac{u_2 - \tfrac{1}{2!} \sig_1^2 \d_\sig^2 u^{\text{\rm BS}}\left(\sig_0\right)}{\d_\sig u^{\text{\rm BS}}\left(\sig_0\right)}, &
\sig_3
    &=  \frac{u_3 - (\sig_2 \sig_1 \d_\sig^2 + \tfrac{1}{3!}\sig_1^3 \d_\sig^3) u^{\text{\rm BS}}\left(\sig_0\right)}{\d_\sig u^{\text{\rm BS}}\left(\sig_0\right)}.
\end{align}
It is important to note that \emph{every term in the sequence $\{ \sig_n \}$ can be computed without integration or special functions}.  To see this, we recall the classical formula $\d_\sig u^\BS(\sig_0) = (T-t)\sig_0(\d_x^2 - \d_x) u^\BS(\sig_0)$ and we note that $u_n = \Lc_n u_0$ is a sum of terms of the form $c_{n,k}(t,T,x,y)\d_x^k(\d_x^2 - \d_x) u^\BS(\sig_0)$ where the coefficients $c_{n,k}(t,T,x,y)$ can be obtained explicitly using \eqref{e32}.  Thus, all of the term in \eqref{eq:bell} can be computed using
\begin{align}
\frac{\d_x^n (\d_x^2 - \d_x ) u^{\BS}(\sig_0)}{(\d_x^2 - \d_x ) u^{\BS}(\sig_0)}
    &= \(\frac{-1}{\sig_0 \sqrt{2(T-t)}}\)^{n} \Hv_{n}(z) , &
z
    &:= \frac{x-k-\frac{1}{2}\sig_0^2 (T-t)}{\sig\sqrt{2(T-t)}} .
\end{align}
where $\Hv_n(z) :=  (-1)^n \ee^{z^2} \d_z^n \ee^{-z^2}$ is the $n$-th Hermite polynomial.

%
%

\section{Examples}
\label{sec:examples}
In this section, we illustrate the flexibility and accuracy of our methodology by applying it to three models: the time-dependent Heston model, the three-halves stochastic volatility model, and the jump-to-default CEV model.  Throughout this section, we always set $(\xb,\yb)=(X_t,Y_t)$, the time-$t$ value of the process $(X,Y)$.


\subsection{Time-dependent Heston model}\label{subset:timedepHest}
We consider the Heston model $(S,Z)$ where $Z$ follows a CIR process with time-dependent mean
$\theta(t)$, vol of vol $\del(t)$ and correlation $\rho(t)$. In $\log$ coordinates $(X,Y)=(\log S,
\log Z)$ we have the following dynamics
\begin{align}
\dd X_t
    &=  -\frac{1}{2} \ee^{Y_t} \dd t + \ee^{\tfrac{1}{2}Y_t} \dd W_t , &
X_t
    &=  x , \\
\dd Y_t
    &=  \( (\kappa\, \theta(t) -\tfrac{1}{2}\del^2(t)) \ee^{-Y_t} - \kappa \) \dd t + \del(t) \, \ee^{-\tfrac{1}{2}Y_t} \dd B_t , &
Y_t
    &=  y  , \label{eq:model.Heston} \\
\dd\<W,B\>_t
    &=  \rho(t) \, \dd t .
\end{align}
From the above dynamics, we obtain
\begin{align}
a(y)
    &= \frac{1}{2} \ee^{y}, &
b(t,y)
    &=  \frac{1}{2} \del^2(t) \ee^{-y}, &
c(t)
    &=  \rho(t) \, \del(t) , &
\alpha(t,y)
    &=  \( (\kappa\, \theta(t) -\tfrac{1}{2}\del^2(t)) \ee^{-y} - \kappa \) .
\end{align}
We choose a particularly simple parameterization of the time-dependent parameters
\begin{align} \del(t)
    &=  \sqrt{\del_0 + \del_1 t} , &
\theta(t)
    &=  \theta_0 + \theta_1 t , &
\rho(t)
    &=  \frac{1}{\del(t)}\(\rho_0 + \rho_1 t\).
\end{align}
This parameterization is by no means required, but it is convenient as $b(t,y)$, $c(t)$ and
$\alpha(t,y)$ acquire an affine dependence in $t$.  Any choice for which $b(t,y)$, $c(t)$ and
$\alpha(t,y)$ can be integrated explicitly with respect to $t$ would be equally tractable.  Using
the results from Sections \ref{sec:approximating} and \ref{sec:impvol} we obtain the following
first order implied volatility approximation
\begin{align}
\sig_0
    &=  \ee^{y/2}, \\
\sig_1
    &=  \frac{3 \rho_0 + (2 t + T) \rho_1}{12 \sig_0}(k-x)
                + \frac{T-t}{24 \sig_0} \Big( -3 \del_0 +6 \theta_0 \kappa +3 \ee^y (-2 \kappa + \rho_0)
                    - (2 t+T) \( \del_1- 2 \theta_1 \kappa -\ee^y \rho_1 \) \Big) .
\end{align}
The second order term $\sig_2$, which we omit for brevity, is quadratic in $(k-x)$.  We
recall that approximations for the time-dependent Heston model were proposed by
\cite{benhamou2010time}.
\par
\begin{figure}
\centering
\begin{tabular}{ccc}
$T=0.125$ & {\qquad} & $T=0.25$ \\
\includegraphics[width=.4\textwidth,height=.2\textheight]{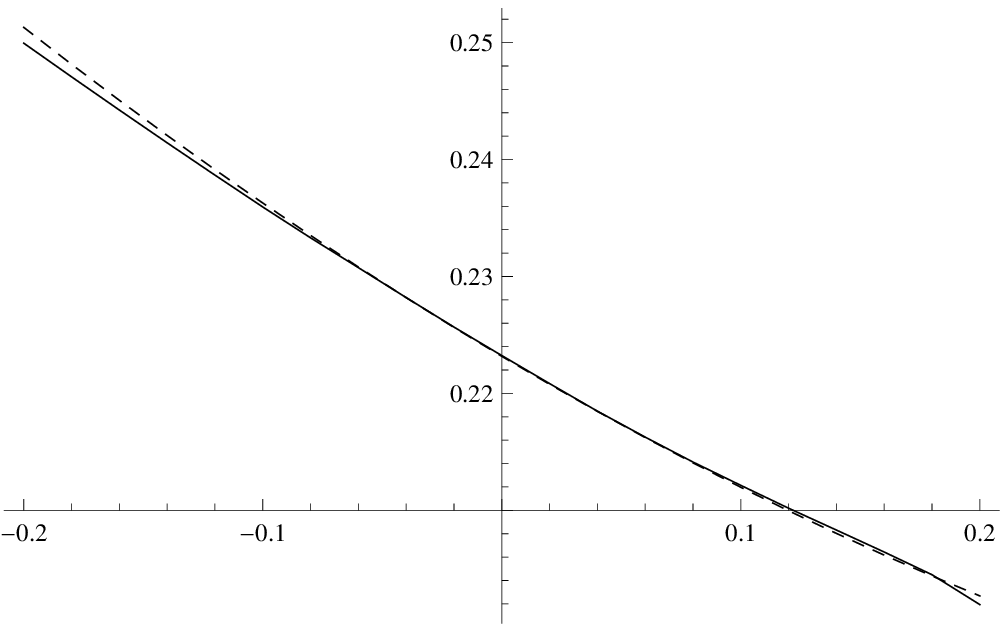} & {\qquad} &
\includegraphics[width=.4\textwidth,height=.2\textheight]{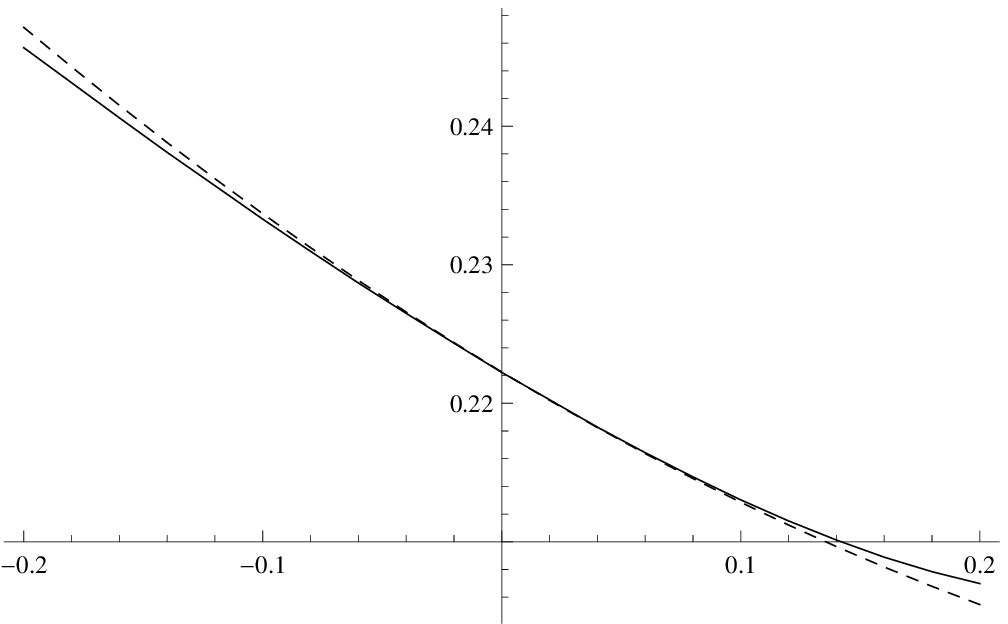} \\
\end{tabular}
\caption{The implied volatility obtained by Monte Carlo (solid) and our second order approximation (dashed) are plotted as a function of $(k-x)$ for the time-dependent Heston model.  Parameters: $t=0$, $\kappa = 1.15$, $\theta_0=0.06$, $\theta_1=-0.08$, $\del_0=0.0625$, $\del_1=-0.16$, $\rho_0=-0.125$, $\rho_1=0.32$, $\ee^y=0.05$.}
\label{fig:heston}
\end{figure}

In Figure \ref{fig:heston} we plot second order implied volatility expansion for two maturities:
$T=0.125$ and $T=0.25$ years.  For comparison we also compute option prices by Monte Carlo
simulation and invert numerically to obtain the corresponding implied volatilities (there is no
exact formula for option prices in the time-dependent Heston model). The mean values of
the parameters $\delta(t)$, $\theta(t)$ and $\rho(t)$ on the interval $[0,0.25]$ and the fixed values $y$ and $\kappa$ correspond to the fixed values of ($\del$, $\theta$, $\rho$, $y$, $\kappa$) used in \cite{forde-jacquier-lee}.


\subsection{Three-halves stochastic volatility model}\label{subset:three-halves}
In the three-halves stochastic volatility model, the stochastic variance process $Z$ satisfies
\begin{align}
\dd Z_t
    &=  \kappa Z_t ( \theta - Z_t ) \dd t + \del Z_t^{3/2} \dd B_t .
\end{align}
In $\log$ coordinates $(X,Y)=(\log S, \log Z)$ we have the following dynamics
\begin{align}
\dd X_t
    &=  -\frac{1}{2} \ee^{Y_t} \dd t + \ee^{\tfrac{1}{2}Y_t} \dd W_t , &
X_t
    &=  x , \\
\dd Y_t
    &=  \Big( \kappa (\theta - \ee^{Y_t} ) - \frac{1}{2}\del^2 \ee^{Y_t} \Big) \dd t
                + \del \, \ee^{\tfrac{1}{2}Y_t} \dd B_t , &
Y_t
    &=  y , \label{eq:model.three-halves} \\
\dd\<W,B\>_t
    &=  \rho \, \dd t .
\end{align}
Thus, we identify
\begin{align}
a(y)
    &= \frac{1}{2} \ee^{y}, &
b(y)
    &=  \frac{1}{2} \del^2 \ee^{y}, &
c(y)
    &=  \rho \, \del \, \ee^{y} , &
\alpha(y)
    &=  \kappa (\theta - \ee^y) - \frac{1}{2}\del^2 \ee^y .
\end{align}
Using the results from Sections \ref{sec:approximating} and \ref{sec:impvol} we obtain the following first order implied volatility approximation
\begin{align}
\sig_0
  &= \ee^{y/2}, &
\sig_1
    &=  -\frac{1}{8} \ee^{y/2} \tau \(-2 \theta  \kappa + \ee^y \(\del^2+2 \kappa -\del  \rho \) \)
            + \frac{1}{4} \ee^{y/2} \delta \rho (k-x), &
\tau
    &:= T-t ,
\end{align}
The second and third order terms $\sig_2$ and $\sig_3$, which we omit for brevity, are both quadratic in $(k-x)$.  To our knowledge, no other implied volatility expansion for the three-halves model appears in literature.
\par
\begin{figure}
\centering
\begin{tabular}{ccc}
$T=0.125$ & {\qquad} & $T=0.25$ \\
\includegraphics[width=.4\textwidth,height=.2\textheight]{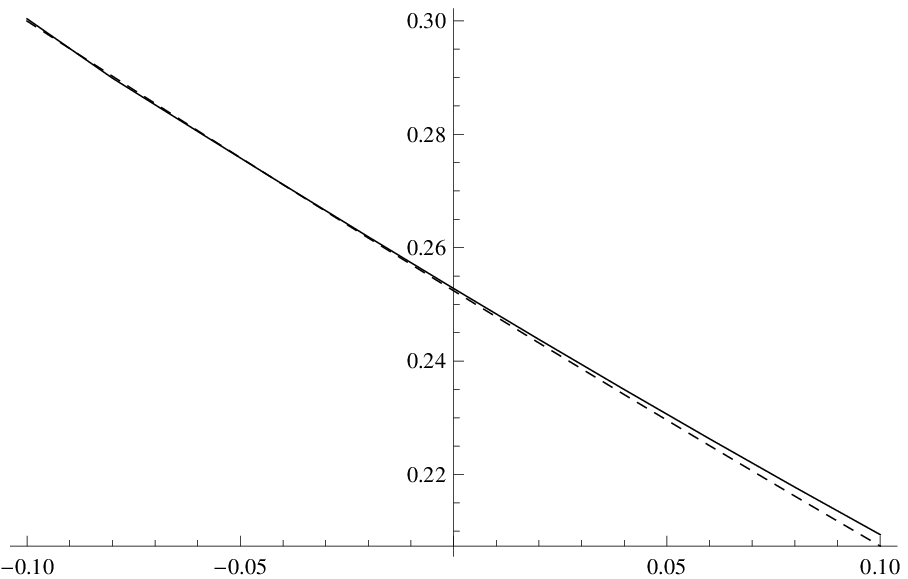} & {\qquad} &
\includegraphics[width=.4\textwidth,height=.2\textheight]{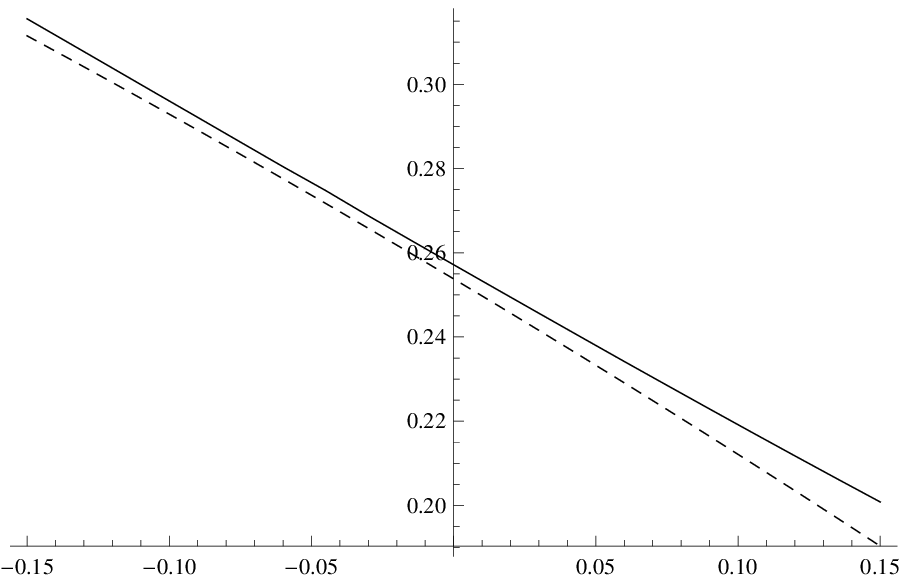} \\
\end{tabular}
\caption{The exact implied volatility (solid) and our third order approximation (dashed) are
plotted as a function of $(k-x)$ for the three-halves model.  Parameters: $\kappa=22.84$,
$\theta=0.4669^2$, $\del=8.56$, $\rho = -0.99$, $\ee^y = 0.245^2$.} \label{fig:three-halves}
\end{figure}

In Figure \ref{fig:three-halves} we plot our third order implied volatility approximation as
well as the exact implied volatility, which we obtain by computing the exact Call price (given in, eg, Proposition 2.2 of \cite{Drimus}) and inverting Black-Scholes numerically. We
use the parameters obtained by \cite{Drimus} calibrating the model to S\&P500 options. Note that
the exact Call price is extremely computationally expensive, as it involves a confluent
hypergeometric function.  By comparison, it is orders of magnitude faster to compute approximate
prices by inserting our implied volatility expansion into the Black-Scholes formula.


\subsection{JDCEV}\label{subset:JDCEV}
\begin{figure}
\centering
\begin{tabular}{ccc}
$\ee^x=1.0$ & {\qquad} & $\ee^x=0.5$ \\
\includegraphics[width=.4\textwidth,height=.2\textheight]{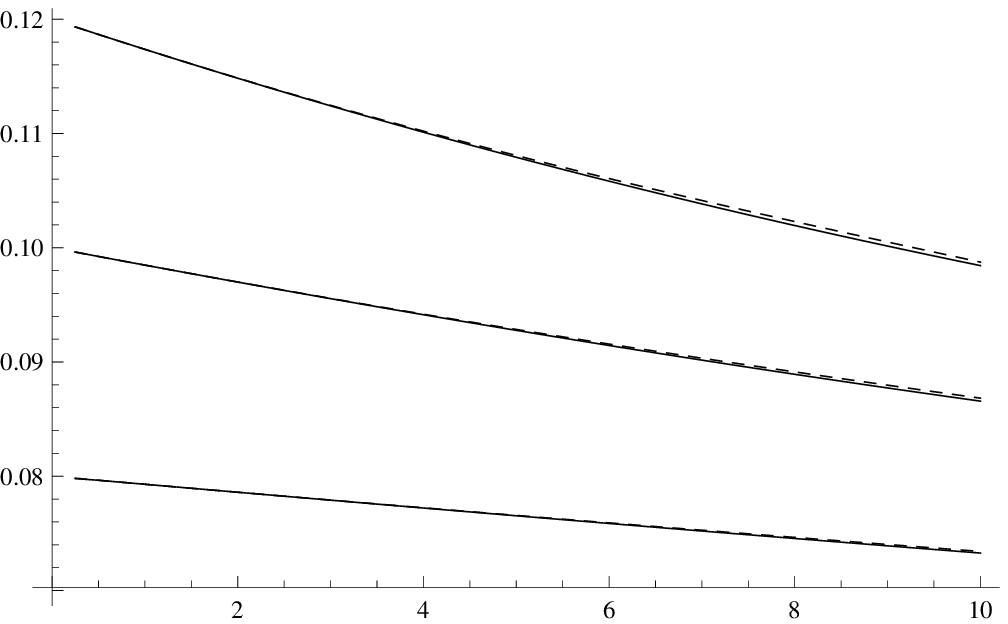} & {\qquad} &
\includegraphics[width=.4\textwidth,height=.2\textheight]{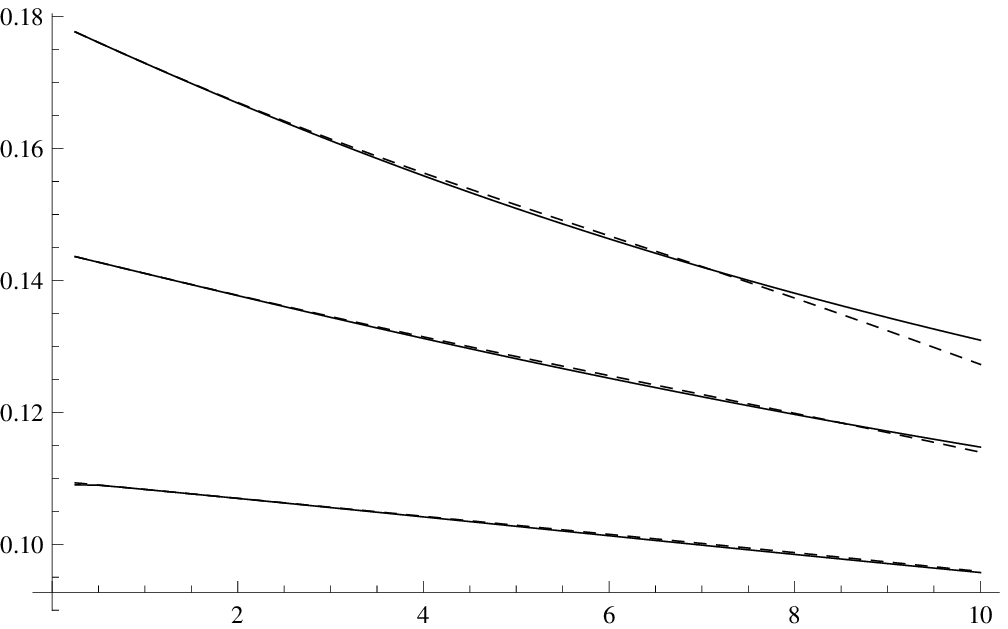} \\
\end{tabular}
\caption{The exact yield curve (solid) and our fourth order approximation (dashed) are plotted as a function of maturity.  Parameters: $\beta=-0.4$, $b=0.04$, top $c=2.0$, middle $c=1.5$, bottom $c=1.0$.}
\label{fig:jdcev}
\end{figure}

As in \cite{JDCEV}, we consider the jump-to-default CEV model, in which an underlying $S_t = \Ib_{\{ \zeta > t\}} \ee^{ X_t }$ has diffusion and killing coefficients $\sig(x) = \del \ee^{\beta x}$ and $\gam(x) = b + c \sig^2(x)$.  Thus we have
\begin{align}
a(x)
    &=  \frac{1}{2} \del^2 \ee^{2 \beta x}, &
\gam(x)
    &=  b + c \, \del^2 \ee^{2 \beta x}, &
\alpha
        &=  \beta = c = 0 .
\end{align}
The yield $Y(t,x;T)$ on a corporate bond that pays $h=1$ at time $T$ if there is no default on the interval $[0,T]$ is given by
\begin{align}
Y(t,x;T)
    &=  \frac{-1}{T-t} \log u(t,x) , &
u(t,x)
    &=  \Eb[ \Ib_{\{ \zeta > T\} } | X_t = x ]
    =       \Eb \[ \ee^{- \int_t^T \gam(X_s) \dd s} | X_t = x \] .
\end{align}
Note that with zero interest rates, the yield corresponds to the credit spread.  Using the results of Section \ref{sec:approximating} we compute
\begin{align}
u_0(t,x)
    &=  \ee^{-\(b+\del^2 c \ee^{2 x \beta }\) \tau}, &
u_1(t,x)
    &=  \ee^{-\(b+\del^2 c \ee^{2 x \beta }\) \tau}
            \Big( -\del^2 b c \ee^{2 x \beta } \tau^2 \beta
            + \frac{1}{2} \del^4 c \ee^{4 x \beta } \tau^2 \beta -\del^4 c^2 \ee^{4 x \beta } \tau^2 \beta \Big) ,
\end{align}
where $\tau = T-t$.  Again, for brevity, we omit higher order terms.  The exact price $u(t,x)$, which requires a  Kummer confluent hypergeometric function, is given in equation (8.13) of \citet{carr}.  In Figure \ref{fig:jdcev} we plot our third order approximation of the yield curve, and the exact yield curve for a variety of model parameters.

%
%

\section{Conclusions and future work}
In this article, we have illustrated how to obtain fast and accurate pricing and implied volatility approximations in a defaultable LSV setting by expanding the drift, diffusion and killing coefficients as a Taylor series.  The resulting price approximations require only a normal CDF.  The resulting implied volatility expansions are explicit.
\par
Mathematica notebooks for computing implied volatilities are provided free of charge on the authors websites (listed below).  Presently, there are notebooks for five well-known models (CEV, Quadratic local volatility, Heston, three-halves stochastic volatility, and SABR).  The websites are updated often, and there are plans to add implied volatility notebooks for models with time-dependent parameters.  Requests for additional models will be entertained as the authors' time permits.
\begin{verbatim}
http://explicitsolutions.wordpress.com
www.princeton.edu/~mlorig
www.math.unipd.it/~stefanop
www.dm.unibo.it/~pascucci
\end{verbatim}


\appendix

%
%

\section{Gaussian derivatives}
\label{sec:gaussian}
Let $\Gam_0=\Gam_0(t,x,y;s,\xi,\om)$ be the Gaussian function in \eqref{e22}. A direct computation reveals
\begin{align}
&\p_{\xi}^{n}\p_{\om}^{m}\left((\xi-\xb)^{h}(\om-\yb)^{k}\Gam_0(t,x,y;s,\xi,\om)\right) \\
&=  (-1)^{n+m} \( \Mc_1^{(x,y)}(t,s) \)^{h} \( \Mc_2^{(x,y)}(t,s) \)^{k}\p_{x}^{n}\p_{y}^{m}\Gam_0(t,x,y;s,\xi,\om), \label{e16}
\end{align}
where $\Mc_1^{(x,y)}(t,s)$ and $\Mc_2^{(x,y)}(t,s)$ are defined in \eqref{eq:M1} and \eqref{eq:M2}.
Now it is quite easy to find the $n$-th order approximation $u_{n}$.
Indeed, recalling that $\At$ denotes the adjoint operator of $\Ac$, we have
\begin{align}
&\At_{1}^{(\xi,\omega)}(s)\Gamma_0(t,x,y;s,\xi,\omega) \\
    &=  \(\At_{1,0}^{(\xi,\omega)}(s)(\xi-\bar{x}) + \At_{0,1}^{(\xi,\omega)}(s)(\omega-\bar{y})\)
            \Gamma_0(t,x,y;s,\xi,\omega) &
            &\text{(by definition of $\At_{1}$)} \\
    &=  \(\Mc_1^{(x,y)}(t,s) \Ac_{1,0}^{(x,y)}(s)+\Mc_2^{(x,y)}(t,s)\Ac_{0,1}^{(x,y)}(s)\)\Gamma_0(t,x,y;s,\xi,\omega) &
            &\text{(by \eqref{e16})} \\
    &= \Gc_1^{(x,y)}(t,s) \Gamma_0(t,x,y;s,\xi,\omega) . \label{eq:identity}
\end{align}
An induction argument proves equation \eqref{e32}.

%
%

\bibliographystyle{chicago}
\bibliography{LPP-bib}

\begin{thebibliography}{}

\bibitem[\protect\citeauthoryear{Benhamou, Gobet, and Miri}{Benhamou
  et~al.}{2010}]{benhamou2010time}
Benhamou, E., E.~Gobet, and M.~Miri (2010).
\newblock Time dependent {H}eston model.
\newblock {\em SIAM Journal on Financial Mathematics\/}~{\em 1\/}(1), 289--325.

\bibitem[\protect\citeauthoryear{Carr and Linetsky}{Carr and
  Linetsky}{2006}]{JDCEV}
Carr, P. and V.~Linetsky (2006).
\newblock A jump to default extended {CEV} model: An application of {B}essel
  processes.
\newblock {\em Finance and Stochastics\/}~{\em 10\/}(3), 303--330.

\bibitem[\protect\citeauthoryear{Drimus}{Drimus}{2012}]{Drimus}
Drimus, G.~G. (2012).
\newblock Options on realized variance by transform methods: a non-affine
  stochastic volatility model.
\newblock {\em Quant. Finance\/}~{\em 12\/}(11), 1679--1694.

\bibitem[\protect\citeauthoryear{Forde, Jacquier, and Lee}{Forde
  et~al.}{2012}]{forde-jacquier-lee}
Forde, M., A.~Jacquier, and R.~Lee (2012).
\newblock The small-time smile and term structure of implied volatility under
  the {H}eston model.
\newblock {\em SIAM Journal on Financial Mathematics\/}~{\em 3\/}(1), 690--708.

\bibitem[\protect\citeauthoryear{Jeanblanc, Yor, and Chesney}{Jeanblanc
  et~al.}{2009}]{yorbook}
Jeanblanc, M., M.~Yor, and M.~Chesney (2009).
\newblock {\em Mathematical methods for financial markets}.
\newblock Springer Verlag.

\bibitem[\protect\citeauthoryear{Lorig}{Lorig}{2013}]{lorigCEV}
Lorig, M. (2013).
\newblock The exact smile of certain local volatility models.
\newblock {\em Quantitative Finance\/}~{\em 13\/}(6), 897--905.

\bibitem[\protect\citeauthoryear{Lorig, Pagliarani, and Pascucci}{Lorig
  et~al.}{2013a}]{lorig-pagliarani-pascucci-1}
Lorig, M., S.~Pagliarani, and A.~Pascucci (2013a).
\newblock A family of density expansions for {L}{\'e}vy-type processes with
  default.
\newblock {\em ArXiv preprint arXiv:1304.1849\/}.

\bibitem[\protect\citeauthoryear{Lorig, Pagliarani, and Pascucci}{Lorig
  et~al.}{2013b}]{lorig-pagliarani-pascucci-2}
Lorig, M., S.~Pagliarani, and A.~Pascucci (2013b).
\newblock Implied vol for any local-stochastic vol model.
\newblock {\em ArXiv preprint arXiv:1306.5447\/}.

\bibitem[\protect\citeauthoryear{Mendoza-Arriaga, Carr, and
  Linetsky}{Mendoza-Arriaga et~al.}{2010}]{carr}
Mendoza-Arriaga, R., P.~Carr, and V.~Linetsky (2010).
\newblock Time-changed {M}arkov processes in unified credit-equity modeling.
\newblock {\em Mathematical Finance\/}~{\em 20}, 527--569.

\bibitem[\protect\citeauthoryear{Pagliarani, Pascucci, and Riga}{Pagliarani
  et~al.}{2013}]{pascucci}
Pagliarani, S., A.~Pascucci, and C.~Riga (2013).
\newblock Adjoint expansions in local {L}\'evy models.
\newblock {\em SIAM J. Financial Math.\/}~{\em 4}, 265--296.

\bibitem[\protect\citeauthoryear{Pascucci}{Pascucci}{2011}]{Pascucci2011}
Pascucci, A. (2011).
\newblock {\em {PDE} and martingale methods in option pricing}.
\newblock Bocconi\&Springer Series. New York: Springer-Verlag.

\end{thebibliography}

%
%

\clearpage

\end{document}